\documentclass[lettersize,journal]{IEEEtran}

\usepackage{url}
\usepackage{soul}
\usepackage{tcolorbox}
\usepackage{listings}
\usepackage{xcolor}
\tcbuselibrary{listings}
\tcbuselibrary{minted}
\usepackage{booktabs}
\usepackage{graphicx}
\usepackage{amsmath}
\usepackage{flushend}
\usepackage{balance}
\usepackage{fancyhdr}
\usepackage{multirow}
\usepackage{hyperref}
\usepackage{enumitem}
\usepackage{pifont}
\usepackage{caption}
\usepackage{svg}
\usepackage{subcaption}

\usepackage{tabularray}
\usepackage{tabularx}
\usepackage{makecell}
\usepackage{array}
\usepackage{float}
\usepackage{threeparttable}

\usepackage[linesnumbered,ruled,vlined]{algorithm2e}

\hypersetup{
    colorlinks=true,
    linkcolor=blue,
    filecolor=magenta,      
    urlcolor=cyan,
    citecolor=blue,
    pdfpagemode=FullScreen,
    }
\usepackage{listings}

\usepackage{xcolor}
\usepackage{diagbox}

\usepackage{longtable}
\usepackage{multirow} 

\definecolor{codegreen}{rgb}{0,0.6,0}
\definecolor{codegray}{rgb}{0.5,0.5,0.5}
\definecolor{codepurple}{rgb}{0.58,0,0.82}
\definecolor{backcolour}{rgb}{0.95,0.95,0.92}

\lstdefinelanguage{TypeScript}{
  keywords={public, private, protected, static, class, interface, extends, implements, constructor, function, return, void, number, boolean, if, else, true, false, null, undefined, throw, new, @Component, @Entry, @Monitor, struct},
  morecomment=[l]{//},
  morecomment=[s]{/*}{*/},
  morestring=[b]",
  morestring=[b]'
}

\lstdefinestyle{customTS}{
    language=TypeScript,
    basicstyle=\ttfamily\small,
    breaklines=true,
    breakatwhitespace=true,
    postbreak=\space,
    commentstyle=\color{gray},
    keywordstyle=\color{blue},
    stringstyle=,
    numbers=none,
    showstringspaces=false,
    tabsize=2,
    frame=single,
    framerule=1pt,
    frameround=tttt,
    backgroundcolor=\color{gray!3},
    basewidth={0.5em,0.5em},
    columns=flexible,
    belowskip=0pt,
    aboveskip=0pt,
    lineskip=-0.5pt,
    xleftmargin=0.02\columnwidth,
    xrightmargin=3pt,
    resetmargins=true,
    moredelim=[is][\color{red}]{@HT@}{@},
    linewidth=0.98\columnwidth,
    breakindent=15pt
}

\lstdefinestyle{customTS_pre}{
    language=TypeScript,
    basicstyle=\ttfamily\small,
    breaklines=true,
    breakatwhitespace=true,
    postbreak=\space,
    commentstyle=\color{gray},
    keywordstyle=\color{blue},
    stringstyle=\color{red},
    numbers=none,
    showstringspaces=false,
    tabsize=2,
    frame=single,
    framerule=1pt,
    frameround=tttt,
    basewidth={0.5em,0.5em},
    columns=flexible,
    belowskip=0pt,
    aboveskip=0pt,
    lineskip=-0.5pt,
    xleftmargin=3pt,
    xrightmargin=3pt,
    resetmargins=true,
    moredelim=[is][\color{red}]{@HT@}{@},
    linewidth=0.5\textwidth,
    breakindent=15pt
}

\lstdefinestyle{customTS_template}{
    language=TypeScript,
    basicstyle=\ttfamily\small,
    breaklines=true,
    breakatwhitespace=true,
    postbreak=\space,
    commentstyle=\color{gray},
    keywordstyle=\color{blue},
    stringstyle=\color{red},
    numbers=none,   
    showstringspaces=false, 
    tabsize=2,    
    frame=single, 
    framerule=1pt, 
    frameround=tttt, 
    backgroundcolor=\color{gray!5},
    basewidth={0.5em,0.5em},
    columns=flexible,
    xleftmargin=0pt, 
    xrightmargin=0pt,
    resetmargins=true,
    linewidth=0.98\columnwidth,
    captionpos=b,      
    framexleftmargin=0.5pt, 
    framexrightmargin=0.5pt, 
    xleftmargin=0.02\columnwidth,
    rulecolor=\color{gray},
    escapeinside={@HT@}{@},
    moredelim=[is][\color{red}]{@HT@}{@},
}

\lstdefinestyle{errorTrace}{
    basicstyle=\ttfamily\small,
    backgroundcolor=\color{black!5},
    frame=single,
    breaklines=true,
    numbers=none,
    xleftmargin=0.5em,
    xrightmargin=0.5em,
    belowskip=0em,
    aboveskip=0em,
    columns=flexible,
    keepspaces=true,
    showstringspaces=false,  
    commentstyle=\color{gray},
    keywordstyle=\color{blue},
    stringstyle=\color{red}, 
    emphstyle=\color{purple},
    morekeywords={Reason, Error, name, message, Stacktrace}, 
}

\usepackage[font=small,labelfont=bf]{caption}

\newcommand{\Finding}[2]{
  \vspace{0.4em}
  \begin{tcolorbox}[
    left=2mm,
    right=2mm,
    left skip=2pt, 
    colframe=gray!80,
    boxrule=0.5pt,
    top=1mm,
    before skip=0mm,
    after skip=0mm
    ]
    \textbf{Finding #1:}
    #2
    \vspace{-0.4em}
  \end{tcolorbox}
}

\usepackage{xspace}
\makeatletter
\DeclareRobustCommand\bmvaOneDot{\futurelet\@let@token\bmv@onedotaux}
\def\bmv@onedotaux{\ifx\@let@token.,\else.\null\fi\xspace}
\DeclareRobustCommand\bmvaTwoDot{\futurelet\@let@token\bmv@twodotaux}
\def\bmv@twodotaux{\ifx\@let@token.,\else.,\null\fi\xspace}
\makeatother

\def\DS{\textbf{ATH Benchmark}\xspace}

\begin{document}
\title{An Empirical Study for GUI Test Migration from Android to OpenHarmony System}

\author{Yakun Zhang, Xinjia Chen, Yiyun Chen, Yuxia Zhang, Mingyi Zhou, Xiang Gao, Shaokun Zhang, Li Li, Yunming Ye

\thanks{Yakun Zhang, Xinjia Chen, Yiyun Chen, Yunming Ye are with Harbin Institute of Technology, Shenzhen, China (email: zhangyk@hit.edu.cn; Schrodinger\_Jia@outlook.com; yiyunchen299@gmail.com; yeyunming@hit.edu.cn)}
\thanks{Yuxia Zhang is with Beijing Institute of Technology ,China (email: yuxiazh@bit.edu.cn)}
\thanks{Mingyi Zhou, Xiang Gao, Li Li are with Beihang University, China (email: zhoumingyi@buaa.edu.cn; xiang\_gao@buaa.edu.cn; lilicoding@ieee.org)}
\thanks{Shaokun Zhang is with Peking University, China (email: skzhang@pku.edu.cn)}}

\markboth{IEEE TRANSACTION ON SOFTWARE ENGINEERING, ~vol.~1, No.~8, April~2026}%
{Shell \MakeLowercase{\textit{et al.}}: An Empirical Study for GUI Test Migration from Android to OpenHarmony System}

\maketitle

\begin{abstract}
With the rapid evolution of mobile operating systems, an increasing number of popular applications originally developed for mature ecosystems such as Android are being migrated to emerging systems  such as OpenHarmony. To reduce the substantial engineering effort required to test the corresponding applications from Android to OpenHarmony, migrating existing GUI test cases
has become a critical problem. However, current research neither proposes solutions tailored for OpenHarmony nor provides a systematic evaluation of migration approaches on this system, leaving developers with limited empirical guidance in practice.

In this paper, we present the first systematic empirical study of test migration from Android to OpenHarmony. Specifically, we first construct a dataset referred to as the Android-to-OpenHarmony Benchmark (\emph{ATH Benchmark}), comprising 36 popular commercial applications with an average of over 9 billion downloads, along with 108 manually designed test cases. Second, we select two state-of-the-art test migration approaches (i.e., ReSPlay and ITeM) and adapt these two approaches on device layer to enable their execution on OpenHarmony. Third, we use the preceding infrastructure to evaluate these two approaches from three perspectives, including \emph{testing performance}, \emph{root causes of failures}, and the \emph{impact of OpenHarmony characteristics}. Our results reveal that existing test migration approaches are less effective (15\% success-rate on ReSPlay and 26\% success-rate on ITeM) in Android-to-OpenHarmony scenarios compared to migration from Android to other systems. Through an in-depth analysis of failed cases, we identify that test performance is primarily hindered by OpenHarmony-specific characteristics, including fundamental technical architecture differences and unique ecosystem traits. Utilizing these findings, we propose an enhanced approach based on ITeM, referred as \emph{ITeM-HM}, which incorporates specific OpenHarmony system features. As a result, ITeM-HM successfully achieves a 214\% success-rate relative improvement over the original ITeM (from 26\% to 81\%). Our dataset and adapted approaches have been open-sourced on \url{https://anonymous.4open.science/r/GUI-Test-Migration/} for future research.



\end{abstract}


\section{Introduction}
\label{sec:introduction}

In recent years, mobile operating systems have continued to evolve, most notably with the emergence of HarmonyOS~\cite{harmonyOS}. HarmonyOS is a commercial operating system built on the OpenHarmony open-source system, integrating with closed-source components~\cite{harmonyOS}. For ease of understanding, we use \emph{OpenHarmony} as a general term to represent this ecosystem in the remainder of the paper. As a next-generation operating system, OpenHarmony aims to establish a unified and multi-device collaborative ecosystem~\cite{harmonyOS}. It has rapidly grown into the world’s third-largest mobile operating system after Android and iOS~\cite{harmonyOS_data}, reaching over 17 million users by the end of 2025~\cite{harmony_user}. Meanwhile, the ecosystem has evolved toward a fully native architecture, with native application frameworks, system services, and execution models~\cite{haptest}.

As the OpenHarmony ecosystem expands, a growing number of applications originally developed for other platforms, especially Android, are being migrated to OpenHarmony. Ensuring the correctness and reliability of these migrated applications requires substantial testing effort. In this context, migrating existing Android GUI test cases to OpenHarmony is becoming increasingly important for both ecosystem builders and application developers.  \ding{182}\emph{From an ecosystem perspective}, migrating existing Android test assets accelerates the population of Huawei App Gallery~\cite{appgallery} while helping maintain application quality at scale. \ding{183}\emph{From a developer perspective}, extending application support to OpenHarmony allows for expansion into a substantial new user base without manual rewriting test cases~\cite{temdroid, macdroid}. Therefore, Android-to-OpenHarmony GUI test migration is not merely a research topic, but an emerging practical need in real-world software development.

Despite the demand, Graphical User Interface (GUI) based test migration from Android to OpenHarmony is not a simple code reuse for three primary reasons. \ding{182}Existing GUI test migration approaches~\cite{craftdroid, testmig, temdroid, apptestmigrator, beyzaei2025automated, RIDA} are typically built upon system-specific execution environments, APIs, and testing infrastructures. Thus, these approaches either focus on Android systems~\cite{craftdroid, temdroid} or investigate cross-system testing between Android and iOS~\cite{testmig}. \ding{183}Although a growing number of researchers have begun to investigate OpenHarmony-related topics~\cite{he2025hacmony, chen2024model, hmtest, haprepair, haptest, typescript2arkts, chen2025arkanalyzerstaticanalysisframework}, no prior work has systematically investigated GUI test migration from Android to OpenHarmony.
\ding{184}More importantly, due to fundamental differences between OpenHarmony and mature platforms, such as application frameworks and execution semantics, existing test migration approaches cannot be directly applied with reliable effectiveness. \textbf{This gap leaves practitioners without empirical evidence on whether current approaches work in this setting, why they fail, and how to design approaches for the system.}

To fill this gap, we present the first empirical study of GUI test migration from Android to OpenHarmony by building a research infrastructure through three aspects. \ding{182}We construct the Android-to-OpenHarmony Benchmark (\textbf{ATH Benchmark}), which contains 108 ground-truth test cases manually designed by four experienced volunteers. These test cases are derived from 36 prominent commercial applications, forming 18 cross-system pairs across six representative categories (e.g., Browser and Map), with an average of over 9 billion downloads form the Google Play Store~\cite{googleplay} and Huawei App Gallery~\cite{appgallery}. \ding{183}We select two state-of-the-art migration approaches: ITeM~\cite{ITeM}, representing LLM-based migration, and ReSPlay~\cite{ReSPlay}, representing matching-based migration. \ding{184} We adapt a device layer to enable execution on OpenHarmony without changing the core migration logic of the approaches, thereby enabling a fair empirical investigation of how existing approaches behave in this new platform setting. Specifically, this layer translates Android-specific device interaction commands into the OpenHarmony system, allowing the ReSPlay and ITeM to obtain device information (e.g., screenshots and layout files) and execute operations (e.g., click and input) on OpenHarmony.

Based on the preceding infrastructure, we first conduct a systematic study for ITeM and Resplay from three perspectives: \emph{migration effectiveness and efficiency}, \emph{root causes of failures}, and \emph{impact of OpenHarmony-specific characteristics}. Specifically, RQ1 evaluates the overall performances of existing approaches. Building on this, RQ2 analyzes the failures identified in RQ1 to explore approach-level migration constraints. Furthermore, RQ3 traces these failures to system-level factors by identifying specific OpenHarmony characteristics that diverge from Android, revealing how these system-specific characteristics affect GUI test migration. The three research questions and key findings are summarized as follows:

\begin{itemize}[leftmargin=1.5em, topsep=0.2em]
\vspace{-0.2em}
  \item \textbf{RQ1 (effectiveness and efficiency): How effective and efficient are existing GUI test migration approaches when migrating test cases from Android to OpenHarmony?} 
We find that both ITeM and ReSPlay suffer substantial performance degradation in the Android-to-OpenHarmony setting. For example, the success-rate of ITeM drops from 54\% to 26\%, and that of ReSPlay drops from 32\% to 15\%. Meanwhile, the execution cost increases sharply, with ITeM taking 12 times longer per step (from 7 seconds to 85 seconds) and ReSPlay costing 3.5 times longer per step (from 29 seconds to 103 seconds). These results suggest that lightweight device layer adaptation is insufficient for migration to OpenHarmony. 

\item \textbf{RQ2 (root causes of failures): What causes existing GUI test migration approaches to fail on OpenHarmony applications?} We find that migration failures are primarily caused by a combination of application-level (28\%), widget-level (16\%) and program-level (56\%) factors. These failures indicate that the challenge is not limited to engineering adaptation, but requires rethinking Android-oriented migration logic to better match the structural characteristics of OpenHarmony applications.
  \item \textbf{RQ3 (impact of OpenHarmony characteristics): How do OpenHarmony-specific characteristics affect GUI test migration?} Our analysis shows that the failures are primarily rooted in technical and ecosystem factors. On the technical side (46\%), the declarative ArkUI framework, the Stage Model, and service-oriented custom widget design reflect a \textit{holistic design philosophy}  (embodied in (1) integral UI design with hierarchy rich in structural rather than just textual and visual information and (2) compositional widget design with a custom widget group rather than an isolated predefined widget that carries specific service functionality) and substantially affect widget understanding and interaction reasoning. On the ecosystem side (26\%), the relative immaturity of the system, rapid evolution, and limited community resources further hinder the effectiveness of existing migration algorithms. Together, these factors fundamentally challenge the direct transfer of Android-oriented solutions.
\end{itemize}
 
Moreover, motivated by the preceding findings, we select ITeM for enhancement as it exhibits superior effectiveness compared to ReSPlay, yet remains constrained by OpenHarmony-specific characteristics. Specifically, we design an OpenHarmony-oriented approach, called \textbf{ITeM-HM}. The key enhancement of ITeM-HM is the introduction of \emph{functional widget groups}, which consist of related widgets that together represent a single functional unit based on OpenHarmony’s holistic design philosophy. To evaluate the performance of ITeM-HM, we further formulate two research questions:

\begin{itemize}[leftmargin=1.5em, topsep=0.2em]
    \item \textbf{RQ4 (Performance of ITeM-HM): How effective is ITeM-HM when migrating test cases from Android to OpenHarmony?} ITeM-HM significantly improves migration effectiveness, increasing the success rate from 26\% (original ITeM) to \textbf{81\%}, corresponding to a \textbf{214\%} improvement. This remarkable enhancement primarily benefits from designing the \emph{functional widget groups} based on the holistic design philosophy of OpenHarmony.
    \item \textbf{RQ5 (ablation study): How much does each component of ITeM-HM contribute to the improvement?} We conduct an ablation study of each enhancement of ITeM-HM, which demonstrates that all components contribute to the improvement and the major improvement stems from the GUI content extraction enhancement (48\%).
\end{itemize}
 
In summary, this paper makes the following contributions:

\begin{itemize}[leftmargin=1.5em, topsep=0.2em]
    \item We present the first empirical study of Android-to-OpenHarmony test migration, providing the first systematic evidence on this important yet underexplored problem.
    \item We construct \textbf{ATH Benchmark}, a benchmark dataset consisting of 108 manually designed test cases for 36 prominent commercial applications with an average downloads of over 9 billion, to support empirical research on Android-to-OpenHarmony GUI test migration.
    \item We adapt and systematically evaluate two state-of-the-art test migration approaches (i.e., ITeM and ReSPlay) on OpenHarmony. Furthermore, we design an OpenHarmony-oriented approach, named \textit{ITeM-HM}, which improves the migration success-rate from 26\% to 81\% by designing the functional widget groups based on the holistic design philosophy of OpenHarmony.
    \item We provide practical insights on GUI test migrations for Android-to-OpenHarmony, identifying general and OpenHarmony-specific practices to improve testing effectiveness in real-world settings. \textbf{Our dataset and the designed approaches are open-sourced on \url{https://anonymous.4open.science/r/GUI-Test-Migration/} ~\cite{open-source} for future research.}
    
\end{itemize}

\section{Background}

\subsection{GUI Test Migration}

\emph{GUI test migration} aims to reduce the manual effort of test development and maintenance by reusing existing GUI test cases across different applications, versions, or systems~\cite{MAPIT, gao2024rule, ramallahliterature}. Instead of creating test cases from scratch for each target context, migration techniques attempt to transfer test logic from a source application or system to a target one, thereby improving the scalability of GUI testing.
In general, prior studies mainly focus on two migration scenarios, which are across-application migration and across-system migration.

\textbf{Across-application migration.} This scenario focuses on migrating test cases between different applications with similar functionalities. Early approaches mainly adopt \emph{matching-based} strategies~\cite{temdroid,craftdroid}, which align widgets across applications using structural, textual, or layout features. Since these approaches generally rely on the assumption of stable widget hierarchies, their effectiveness degrades when widget structures or semantics diverge significantly. To address these limitations, recent studies have proposed \textit{LLM-based} approaches~\cite{beyzaei2025automated, li2025reusedroid, apptestmigrator}. Representative approaches, such as ITeM~\cite{ITeM}, infer test intentions and use semantic reasoning to identify semantically equivalent interaction widgets, making migration more flexible under diverse application designs.

\textbf{Across-system migration.} This scenario targets migrating test cases of the same application across different operating systems, such as Android and iOS~\cite{ji2023vision, gao2024rule, behrang2019test}, Android and Web~\cite{lin2022gui}. In this setting, migration approaches typically rely on the correspondence of widget attributes, such as view hierarchies, textual labels, and visual appearance. Representative \textit{matching-based} approaches~\cite{khalili2024semantic, lin2022gui, RIDA}, such as ReSPlay~\cite{ReSPlay}, migrate test cases by matching widgets with similar structures, texts, and appearances across systems. These approaches generally assume predictable event handling across systems, so that similar widgets exhibit comparable behaviors during execution.

Despite this progress, research on Android-to-OpenHarmony GUI test migration remains limited. Compared with previous settings, this scenario introduces \textbf{additional challenges} caused by differences in application frameworks, widget exposure mechanisms, event triggering logic, and execution assumptions. These differences substantially increase the difficulty of test reuse and expose the limitations of existing migration approaches in this emerging system. Therefore, \emph{there is an urgent need for empirical evaluation and approach proposed for this emerging scenario.}

\subsection{OpenHarmony Application Framework}
\label{sec:HMbackground}
OpenHarmony is an operating system designed for collaborative multi-device scenarios, whose application framework differs from traditional mobile operating systems in terms of overall design philosophy and execution mechanisms. In the following, we briefly introduce the evolution of OpenHarmony, its fully native architecture, and the implications of these differences for GUI test migration.

\textbf{The evolution of OpenHarmony.} Early versions of OpenHarmony supported both native applications and compatibility-based execution of Android applications~\cite{harmonyOS}. However, as the OpenHarmony ecosystem continues to evolve, the system has gradually shifted toward a native architecture. Starting from HarmonyOS 5.0~\cite{harmonyOS_5}, the Android runtime compatibility layer is progressively removed, and OpenHarmony adopts a completely native application framework, system services, and execution model~\cite{harmony_differ}. As a result, these differences become increasingly significant for GUI test migration.

\textbf{Fully native architecture.} OpenHarmony native applications are fully built upon three key elements: the \textit{ArkTS}~\cite{huawei_arkts_guide} programming language, the \textit{ArkUI}~\cite{huawei_arkui_guide} declarative UI framework, and the \textit{Stage Model}~\cite{huawei_arkui_stage} for the application lifecycle and page management~\cite{harmony_release}. 

At the language level, native applications are developed using \textit{ArkTS}, a language derived from \texttt{TypeScript} and optimized for OpenHarmony application development. ArkTS emphasizes the declarative UI construction and reactive state management, encouraging developers to describe the application behaviors in terms of state changes rather than the imperative control flow~\cite{typescript2arkts}.

At the UI level, the native UI framework is ArkUI, a declarative framework that automatically updates UI \textit{components} in response to state changes. These UI components are conceptually similar to Android \textit{widgets}; for clarity, we use the term \textit{widget} throughout this paper. Unlike traditional imperative UI frameworks, ArkUI abstracts away much of the low-level rendering and event dispatch process, which changes how widgets are exposed and observed during testing.

At the execution level, application lifecycle and page navigation are managed through the \textit{Stage Model}. Instead of adopting Android’s activity-centric mechanism, the Stage Model provides a unified abstraction for page organization, widget lifecycle, and resource coordination across devices. This affects page transitions and executions.

\textbf{Impact on GUI test migration.} The above architectural characteristics introduce new challenges for existing GUI test migration approaches. First, under ArkUI’s declarative and state-driven paradigm, UI hierarchies are dynamically constructed and recomposed based on application state, making widget identification less stable than in traditional Android settings. Second, ArkTS-based reactive state management separates user interactions from explicit event-widget bindings, making it difficult to infer interaction semantics and execution feedback during migration. Third, the Stage Model replaces Android’s activity-centric lifecycle with a container-based page and widget management mechanism, which affects how migration approaches determine whether a migrated step has been executed correctly. As a result, existing GUI test migration approaches that use assumptions (e.g., stable widget hierarchies or predictable event handling) are difficult to ensure reliable execution in OpenHarmony environment.

Therefore, although existing test migration approaches have shown promise in previously studied Android-related settings, their underlying migration assumptions (e.g., stable widget hierarchies or predictable event handling) may not hold in OpenHarmony. These architectural and execution differences motivate a dedicated examination of GUI test migration in the Android-to-OpenHarmony scenario.

\section{Methodology}
\label{sec:methodology}

This section presents the methodology of our empirical study on Android-to-OpenHarmony GUI test migration. Specifically, \ding{182}we first construct the Android-to-OpenHarmony Benchmark (\textbf{ATH Benchmark}, see \ref{sec:dataset-construction}), including cross-system application pairs and manually constructed ground-truth test cases. \ding{183}We then introduce the representative migration approaches~\cite{ReSPlay, ITeM} and adapt them to execute in OpenHarmony (see \ref{sec:evaluation-setting}), followed by the research questions, metrics, and experimental configuration.

Note that, compared with prior studies~\cite{craftdroid, beyzaei2025automated, gao2024rule, ji2023vision, ramallahliterature} that mainly focus on Android-only or Android-to-iOS settings, our study emphasizes \emph{industrial realism} in both dataset construction and evaluation. Specifically, \ding{182} the benchmark is derived from prominent commercial applications that are available in both the Google Play Store and the official App Gallery; and \ding{183} the evaluation targets real Android-to-OpenHarmony migration tasks rather than synthetic or simplified cross-system settings. 

\subsection{Dataset Construction}
\label{sec:dataset-construction}
To support a systematic empirical study on Android-to-OpenHarmony GUI test migration, we construct the \textit{Android-to-OpenHarmony Benchmark (\DS)}. The benchmark is designed to provide a reliable experimental foundation for this emerging migration scenario by ensuring that the selected subjects and test cases are representative, executable, and comparable across systems. In the following, we first describe the selection of application subjects and then present the construction of ground-truth functional test cases.

\subsubsection{Subject Selection}
To construct the benchmark, we first collect a set of mobile applications for analyzing Android-to-OpenHarmony test migration. Second, we select three representative functionalities for each application.

\textbf{Application Selection.} Given both the effectiveness and feasibility of the evaluation experiments, the selection of applications followed the four criteria~\cite{yuan2015app, shauvik2015input, am2001gui} below:

\begin{itemize}[leftmargin=*]
    \item \textbf{Cross-System Compatibility}: Each application must be able to execute normally on the Android system. In addition, it should be successfully launched and executed on OpenHarmony, through native execution rather than compatibility mechanisms (e.g., Droitong~\cite{droitong} and EasyAbroad~\cite{easyabroad}).

    \item \textbf{GUI Interaction Complexity}: To avoid trivial migration scenarios, selected applications must feature sophisticated GUI structures. We require subjects to possess multiple hierarchical screens and support diverse user interactions (e.g., asynchronous page transitions, gesture-based scrolling, and complex form inputs) to facilitate the generation of representative test sequences.

    \item \textbf{Domain Diversity}: The dataset must cover diverse application categories to mitigate domain-specific bias and enhance the generalizability of our empirical findings.

    \item \textbf{Application Popularity}: Applications are required to be widely used in practice, measured by their download volumes in official channels, to ensure that the selected subjects reflect the real-world ecosystem.
\end{itemize}

Following existing test migration studies~\cite{craftdroid, fruiter}, we select six common application categories (e.g., Browser, Entertainment). For each category, we choose the top five commercial applications \emph{with an average downloads of over 9 billion} from Google Play~\cite{googleplay} and Huawei App Gallery~\cite{appgallery} based on the above four criteria. This process results in a dataset of 36 representative applications (e.g., Booking~\cite{booking}, Meitu~\cite{meitu}) as our study subjects. Each application is manually verified to ensure that its basic GUI functionalities and interaction workflows operate correctly in both environments. 

Note that, the scale of our benchmark is partly constrained by the current stage of the OpenHarmony ecosystem. As an emerging platform, OpenHarmony still provides only a limited number of large-scale commercial applications that are natively available on both Android and OpenHarmony, making cross-system subject collection substantially more difficult than in mature Android-centric settings. Rather than constructing a toy benchmark from small or simplified applications, we deliberately focus on representative industry applications with real-world popularity and non-trivial GUI workflows. To the best of our knowledge, no prior work has established a dedicated benchmark for Android-to-OpenHarmony GUI test migration, which means that the entire process (including candidate discovery, cross-system screening and verification, and aligned ground-truth test case construction) requires substantial manual effort. Despite these challenges, we finally construct \DS, which contains 36 representative applications, a scale comparable to existing state-of-the-art migration studies such as ReSPlay with 12 applications~\cite{ReSPlay} and FrUITeR with 20 applications~\cite{fruiter}. Table~\ref{tab:app_info} summarizes the statistics of the selected applications in the benchmark.

\textbf{Functionalities Selection.} For each application category, we select three representative functionalities, as shown in Table ~\ref{tab:function}. The selection of these three functionalities is driven by two aspects: (1) \emph{Methodological Alignment}: we follow the experimental designs of CraftDroid~\cite{craftdroid} and RIDA~\cite{RIDA}, which evaluate applications using two or three core usage scenarios (e.g., Browse); (2) \emph{Industrial Relevance}: we prioritize the primary functionalities highlighted in the Google Play Store~\cite{googleplay} and App Gallery~\cite{appgallery} of OpenHarmony descriptions to ensure our test cases reflect the most frequent and critical user workflows in real-world commercial environments. 

\begin{table}[t]
  \small
  \caption{Details of Selected Applications and Test Case Functions}
  \label{tab:app_info}
  \setlength{\tabcolsep}{3pt}
  \vspace{-0.5em}
  \centering
  \resizebox{\columnwidth}{!}{%
    \begin{tabular}{llccl}
    \toprule
    \multicolumn{1}{c}{\multirow{2}{*}{Category}} & \multicolumn{1}{c}{\multirow{2}{*}{App}} & Downloads & Downloads & \multicolumn{1}{c}{\multirow{2}{*}{Functionality}} \\
    &  & (Android) & (OpenHarmony) &  \\
    \toprule
    
    \multirow{3}{*}{Browser} & Alook Browser & 50K+ & 114K+ & Add new tab \\
    & UC Browser & 1B+ & 5M+ & Search \\
    & Baidu & 5B+ & 38M+ & Back \\
    \midrule
    
    \multirow{3}{*}{Entertainment} & Bilibili & 1B+ & 17M+ & Search \\
    & QQ Music & 5M+ & 11M+ & Browse \\
    & Tiktok & 69B+ & 46M+ & Like \\
    \midrule
    
    \multirow{3}{*}{Photography} & Meitu & 100M+ & 5M+ & Image \\
    & Beauty Cam & 50M+ & 2M+ & Video \\
    & Hypic & 1B+ & 3M+ & Collage \\
    \midrule
    
    \multirow{3}{*}{Shopping} & Booking & 500M+ & 295K+ & Search \\
    & JD & 5M+ & 31M+ & Add to cart \\
    & Pinduoduo & 67B+ & 37M+ & Edit cart \\
    \midrule
    
    \multirow{3}{*}{Social Media} & Rednote & 10M+ & 17M+ & Search \\
    & Weibo & 10M+ & 8M+ & Follow \\
    & QQ & 24B+ & 32M+ & Post \\
    \midrule

    \multirow{3}{*}{Map} & Amap & 10M+ & 43M+ & Search \\
    & BaiduMap & 10M+ & 15M+ & Locate \\
    & TencentMap & 3B+ & 3M+ & Navigate \\    
    \bottomrule
\end{tabular}
}
\begin{flushleft}
    \small
    \textit{Note: Each application includes three functionalities for testing.}
\end{flushleft}
\vspace{-1.2em}
\end{table}
\begin{table}[htbp]
  \caption{Functionalities of Test Cases for each Application}
  \small
  \label{tab:function}
  \centering
  \resizebox{\columnwidth}{!}{
  \begin{tabular}{cll}
    \toprule
    \textbf{Category} & \textbf{Functionality} & \textbf{Description}\\
    
    \midrule
    \multirow{3}{*}{Browser}
    &Add new tab & Navigate to switch tab page and click add tab button \\ 
    &Search & Use search bar to search a URL \\ 
    &Back & Click back button to return to the previous page \\

    \midrule
    \multirow{3}{*}{Entertainment}
    &Search & Use search bar to search a media \\ 
    &Browse & Pause, play, and switch media \\
    &Like & Like a media and subscribe the author \\

    \midrule
    \multirow{3}{*}{Photography}
    &Image & Navigate to camera page and take a photo\\ 
    &Video & Navigate to camera page and record a video \\ 
    &Collage & Navigate to collage page and collage two recent photos \\

    \midrule
    \multirow{3}{*}{Shopping}
    &Search & Use search bar to search a product \\ 
    &Add to cart & Add the first search result item to cart\\ 
    &Edit cart & Open cart and remove the first item from cart \\

    \midrule
    \multirow{3}{*}{Social Media}
    &Search & Use search bar to search a news\\ 
    &Follow & Search an account and follow it \\ 
    &Post & Navigate to post page, write and publish a post\\

    \midrule
    \multirow{3}{*}{Map}
    &Search & Use search bar to search a location \\ 
    &Locate & Browse the map and locate current position \\
    &Navigate & Navigate to the first search result location \\
    
  \bottomrule
\end{tabular}
}
\end{table}

\subsubsection{Test Case Construction}
\label{sec:test_case}
After selecting the experiment subjects, we establish a systematic pipeline to construct and validate ground-truth functional test cases, including expert-led development, test cases construction for Android and OpenHarmony and rigorous quality control.  

\textbf{Expert-led Development.} To simulate realistic GUI testing workflows, we recruit four researchers with at least three years of experience in both Android and OpenHarmony development. We adopt the \emph{back-to-back testing} approach~\cite{backtoback}, in which one researcher is responsible for constructing functional test cases, while another independently reviews and validates them until full agreement is reached. All participants possess substantial domain knowledge and hands-on experience, which helps ensure the correctness and reliability of the test assets.

The volunteers decompose each functional test case into a sequence of ordered GUI interaction steps. For example, the \emph{search} functionality in the browser category,  can be decomposed into three steps, including (1) clicking the search input weight, (2) entering a query keyword, and (3) validating the resulting page as an oracle step. Each step is recorded in a unified, structured format that captures the control locator information, interaction actions, execution semantics, and the corresponding GUI state. This representation enables the assets to be effectively reused for cross-system migration analysis.

\begin{figure}[t]
    \centering
    \begin{subfigure}{0.5\textwidth}
        \includegraphics[width=\linewidth]{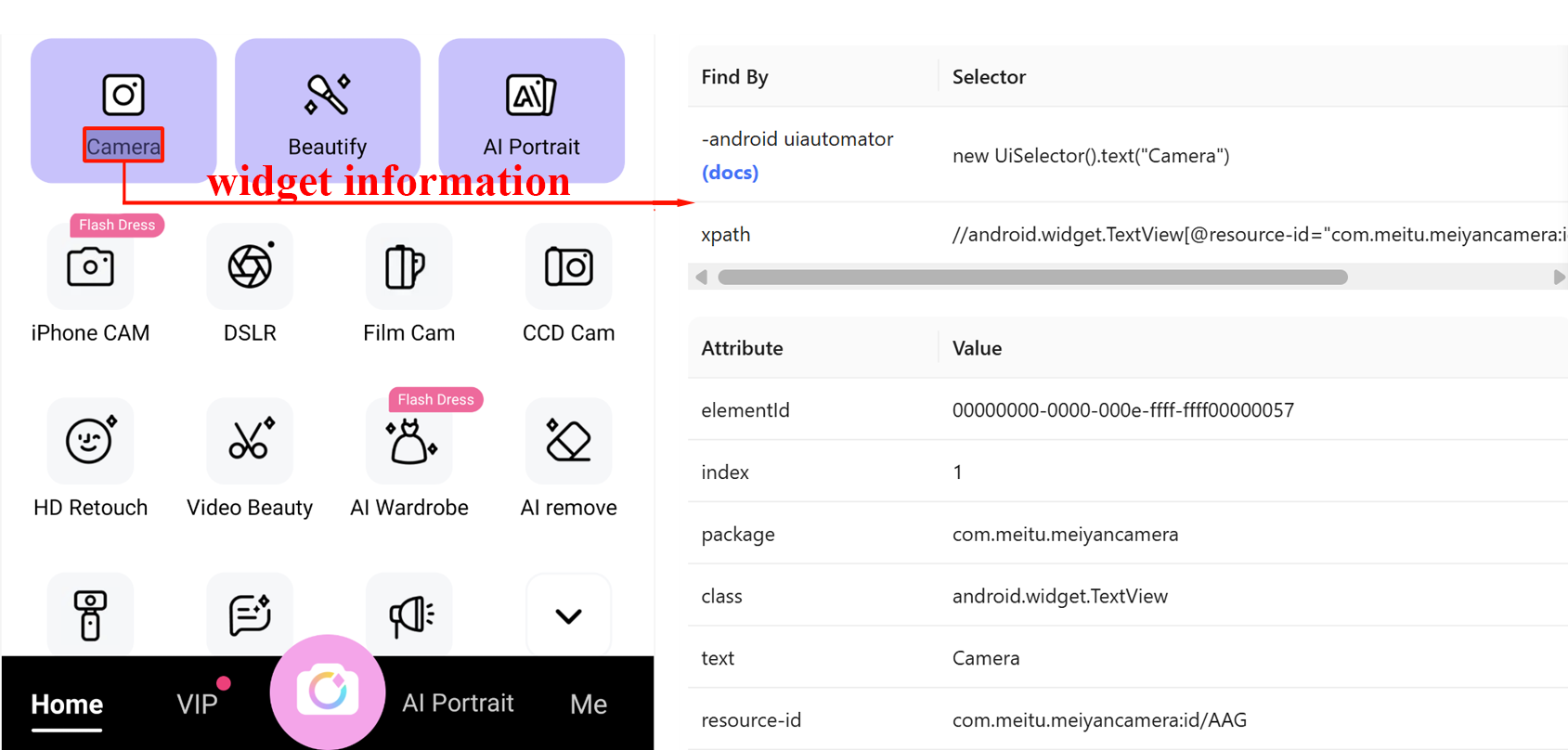}
        \caption{Android system}  
    \end{subfigure}
    \hfill
    \begin{subfigure}{0.4\textwidth}
        \includegraphics[width=\linewidth]{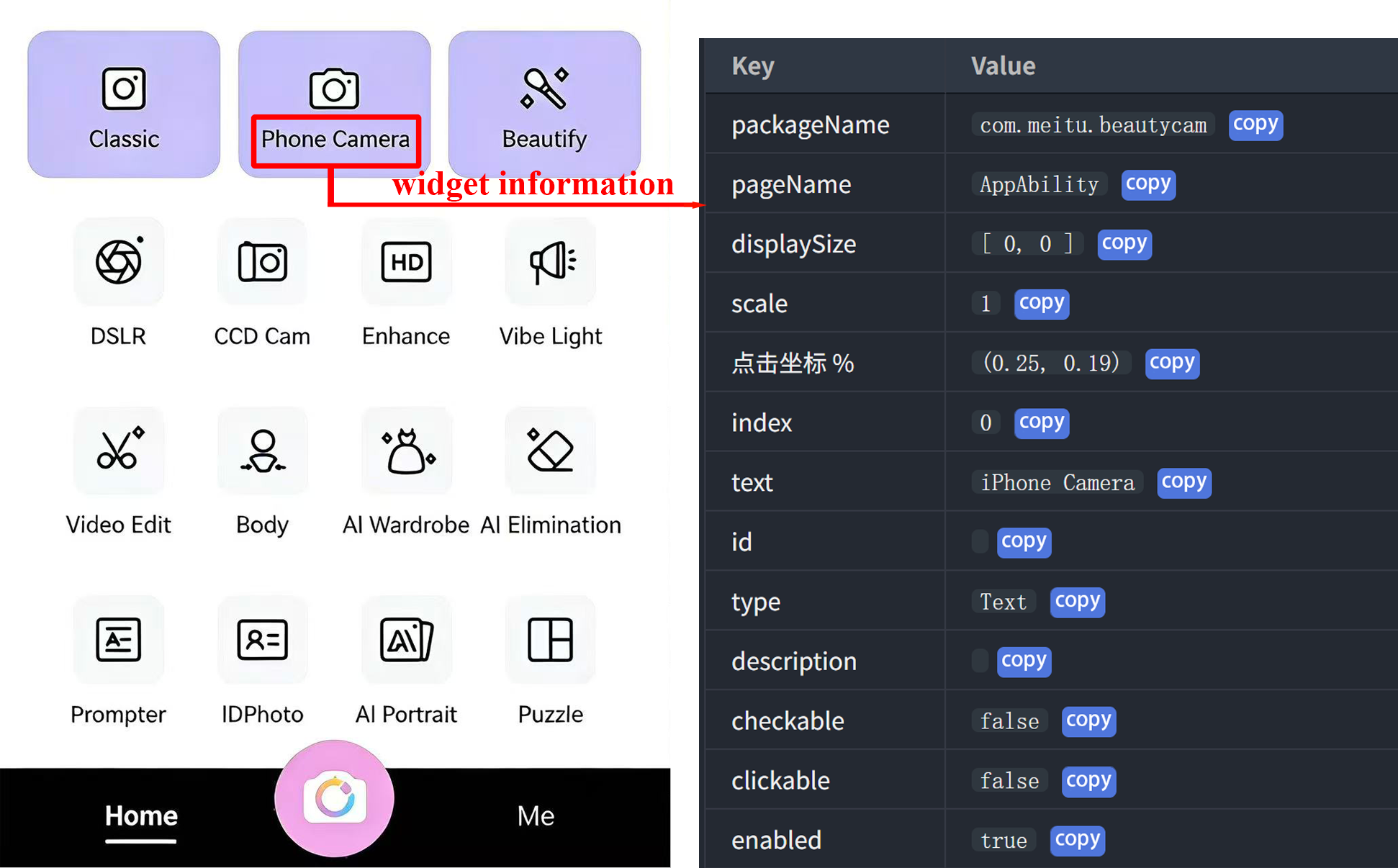}
        \caption{OpenHarmony system}  
    \end{subfigure}
    \vspace{-0.3em}
    \caption{Widget information of BeautyCam on the Android and OpenHarmony system}
    \label{fig:case_example}
\vspace{-1.5em}
\end{figure}

\textbf{Test Cases for Android System.} To establish a robust ground-truth, we construct functional test cases using the Appium framework~\cite{appium} and extracted fine-grained GUI widget metadata, as illustrated in Figure~\ref{fig:case_example}. The construction process follows a four-step workflow. First, for each functional test case, we decompose the workflow according to its functional logic into several ordered steps. Second, we record the information of each test step from Appium, including \texttt{class}, \texttt{resource-id}, \texttt{text}, \texttt{content-desc}, and \texttt{clickable}. Third, we define its \texttt{event\_type}, indicating whether the step is a \emph{GUI event} (user iteration) or an \emph{oracle event} (assertion/verification), and specify the concrete action to be performed. Fourth, we use the adb~\cite{adb_tool} tool to export the current \texttt{Activity} and GUI hierarchy during the execution of each step to record the runtime state. 
Finally, we save a JSON-formatted file describing the test steps on Android.

\textbf{Test Cases for OpenHarmony System.} To ensure cross-system comparability, we use UIViewer~\cite{uiviewer} to extract GUI interface information, depicted in Figure ~\ref{fig:case_example}. A key challenge in this process is the structural disparity between the two operating systems. To mitigate this, we meticulously establish an attribute mapping schema to maintain consistency with the Android baseline. Specifically, we record OpenHarmony attributes and map them to their Android counterparts: \texttt{type} (corresponding to Android \texttt{class}), \texttt{id} (corresponding to \texttt{resource-id}), \texttt{text}, \texttt{description} (corresponding to \texttt{content-desc}), \texttt{clickable}, \texttt{parent\_text}, \texttt{sibling\_text}, \texttt{pageName} (corresponding to \texttt{Activity}), and \texttt{packageName} (corresponding to \texttt{package}). The construction of test steps follows the same procedure as on Android. For each functional test case, we decompose it into multiple test steps and defined event\_type and action for each step. We further use the hdc~\cite{hdc_tool} tool to obtain the current Activity and GUI hierarchy. 
Similar to the Android setup, we save a JSON-formatted file describing the test steps on the OpenHarmony system.

\textbf{Quality Control.} To assess annotation consistency, we use \textit{Fleiss' kappa coefficient}, a widely adopted metric for measuring inter-rater agreement~\cite{measuring}. Under this quality-control process, the resulting kappa value reaches \emph{0.94}, indicating a high degree of consistency and usability. Therefore, the constructed test cases not only have practical value, but also provide a reliable foundation for cross-system migration analysis.

\subsection{Evaluation Setting}
\label{sec:evaluation-setting}
In this section, we describe the evaluation setting of our empirical study. Specifically, we introduce the selected GUI test migration approaches, present the evaluation metrics, report the experimental configuration, and describe the OpenHarmony-oriented adaptation workflow that enables the evaluated approaches to execute in the target environment.

\subsubsection{Migration Approaches}
\label{sec:approaches}
We select two representative GUI test migration approaches for evaluation.

 ITeM~\cite{ITeM} is the state-of-the-art GUI test migration approach for Android applications and represents the migration approaches driven by LLM. Migration is mainly driven by three core components: test intention extraction (to summarize the test intention of each step), GUI content extraction (to filter and represent candidate widgets for LLM) and action execution (to execute LLM's actions in dynamic reasoning).
 
 ReSPlay~\cite{ReSPlay} is a migration-oriented approach that has been validated in Android-to-iOS migration scenarios and represents widget-matching-based migration approaches. 
 
 Note that, we do not select the recently proposed LLMRR~\cite{llmrr} approach in our study, as it relies on manual configuration of playback sequence, which costs substantial human effort and does not align with the low-manual-intervention requirements of large-scale real-world scenarios. 

\subsubsection{Metrics}
\label{sec:metrics}
To comprehensively evaluate the study objects, we define four metrics, grouped into two categories: effectiveness related metrics and execution cost related metrics.

\paragraph{Effectiveness Metrics}
To evaluate the effectiveness of the approaches, we introduce two metrics, namely \textit{success-rate} and \textit{step-accuracy}, following ReSPlay and ITeM~\cite{ReSPlay,ITeM}.

\underline{Success-rate.} This metric is calculated as the ratio of successfully migrated test cases to the total migrated test cases.

\begin{equation}
    \text{success-rate} = \frac{\text{Number of Successfully Migrated Cases}}{\text{Total Number of Cases}} \times 100\%
\end{equation}

\underline{Step-accuracy.} This metric is calculated as the ratio of correctly executed steps to the total action steps across all migrated test cases.

\begin{equation}
\text{step-accuracy} = \frac{\text{Number of Correct Steps}}{\text{Total Number of Steps}} \times 100\%
\end{equation}

\paragraph{Execution Cost}
To evaluate the execution cost of the migrated test cases, we introduce two metrics, namely \textit{time-per-step} and \textit{tokens-per-step}.

\underline{Time-per-step.} This metric measures the average time cost required to complete and confirm a correct interaction step during test execution.

\begin{equation}
    \text{time-per-step} = \frac{\text{Total Execution Time}}{\text{Number of Correct Steps}}
    \end{equation}

\underline{Tokens-per-step.} This metric measures the average number of tokens consumed by the large language model (LLM) to complete and confirm a correct interaction step.

\begin{equation}
    \text{tokens-per-step} = \frac{\text{Total Token Consumption}}{\text{Number of Correct Steps}}
\end{equation}

Since ReSPlay does not rely on an LLM, the \textit{tokens-per-step} metric is only reported for ITeM.

\subsubsection{Experimental Configuration}
The experiments of ITeM are conducted on a Windows PC equipped with a 13th Gen Intel(R) Core(TM) i9-13900H CPU and 16GB RAM. The OpenHarmony applications are executed on a physical device (HUAWEI NOVA 14) equipped with HarmonyOS 5.0.0. The LLM used by ITeM is GPT-5.

The experiments of ReSPlay are conducted on a Linux PC equipped with an Intel(R) Xeon(R) Silver 4310 CPU, 256GB RAM and four GeForce RTX 3090 GPUs. The device to execute the OpenHarmony applications is the same as that of ITeM. ReSPlay does not require LLMs.

\subsubsection{Experimental Workflow}
\label{sec:workflow}

This section explains why existing test migration approaches require adaptation for OpenHarmony and how we achieve it. 

We first analyze that both ITeM and ReSPlay are tightly coupled with the Android execution environment. Specifically, they rely on adb commands and the Appium framework to obtain GUI information from the target device and inject interaction events during migration. Since these mechanisms are designed for Android devices only, they cannot directly support OpenHarmony. \emph{As a result, existing migration approaches cannot be directly applied to the Android-to-OpenHarmony scenario without additional system-level adaptation.}

To bridge this gap, we then design and implement a \textbf{device layer adaptation} that connects existing migration approaches with OpenHarmony devices while preserving their original migration logic. The adaptation is carried out in three aspects. \ding{182}We re-implement the basic device interaction interfaces required by migration approaches using hdc commands, including device connection, screen capture, layout hierarchy extraction, widget localization, application management (e.g., launch, stop, and data clearance), and operation injection (e.g., click, double-click, swipe, keyboard input, and back events). \ding{183}To address the representation gap between Android and OpenHarmony, we normalize screenshots, layout hierarchies, widget attribute names, and basic data types, and expose them in a format compatible with the original implementations based on Appium and uiautomator2~\cite{android-uiautomator}. \ding{184}Based on the above interfaces and normalized representations, we build an execution bridge that allows migration approaches to obtain GUI states from OpenHarmony devices and inject actions into them in a unified manner.

Through this adaptation, the evaluated approaches are able to execute on native OpenHarmony devices under a controlled and comparable setting. \emph{This adaptation layer serves as the engineering foundation of our study, enabling a fair empirical evaluation of existing GUI test migration approaches and supporting the subsequent analysis of migration failures and OpenHarmony-oriented enhancement.}
\section{Evaluation}

In this section, we conduct a comprehensive experimental evaluation to assess the performance of GUI test migration approaches on Android-to-OpenHarmony scenario.

\subsection{RQ1: Effectiveness and Efficiency}
\label{sec:RQ1}

After applying the device layer adaptation described in Section~\ref{sec:workflow}, both ReSPlay and ITeM can execute successfully on OpenHarmony.
To evaluate their performance, we apply them to the ATH Benchmark for test migration tasks. Given the non-determinism of LLMs, we conduct three independent runs on ITEM and adopt the median results. 

\begin{table}[t]
  \centering
  \small
  \caption{Performance Metrics by Category}
  \label{tab:rq12}
  \vspace{-0.5em}
  \resizebox{\columnwidth}{!}{%
  \begin{tabular}{llcccc}
    \toprule
    \textbf{Category} & \textbf{Techniques} & \textbf{Success-rate} & \textbf{Step-accuracy} & \textbf{Time(s/step)} & \textbf{Token(tokens/step)} \\
    \midrule
    \multirow{2}{*}{Browser} 
      & ReSPlay & 11\% & 11\% & 173 & - \\
      & ITeM    & 44\% & 48\% & 47 & 13977 \\
    \midrule
    \multirow{2}{*}{Entertainment} 
      & ReSPlay & 44\% & 52\% & 100 & - \\
      & ITeM    & 11\% & 13\% & 93 & 37775 \\
    \midrule
    \multirow{2}{*}{Photograph} 
      & ReSPlay & 0\% & 16\% & 58 & - \\
      & ITeM    & 11\% & 29\% & 68 & 13649 \\
    \midrule
    \multirow{2}{*}{Shopping} 
      & ReSPlay & 11\% & 18\% & 157 & - \\
      & ITeM    & 33\% & 56\% & 98 & 78854 \\
    \midrule
    \multirow{2}{*}{Social Media} 
      & ReSPlay & 0\% & 11\% & 66 & - \\
      & ITeM    & 22\% & 27\% & 101 & 56601 \\
    \midrule
    \multirow{2}{*}{Map} 
      & ReSPlay & 22\% & 13\% & 79 & - \\
      & ITeM    & 33\% & 74\% & 88 & 40385 \\
    \midrule
    \multirow{2}{*}{Overall} 
      & ReSPlay & 15\% & 18\% & 103 & - \\
      & ITeM    & 26\% & 42\% & 85 & 45780 \\
    \bottomrule
  \end{tabular}
  }
\vspace{-1.2em}
\end{table}

\subsubsection{Effectiveness}
To evaluate the effectiveness of migration, we calculate the \underline{success-rate} and \underline{step-accuracy} of Resplay and ITeM on the ATH Benchmark.

\textbf{Experimental results.} As summarized in Table~\ref{tab:rq12}, both ReSPlay and ITeM exhibit suboptimal performance when migrated to the OpenHarmony ecosystem. Specifically, ReSPlay achieves a \underline{success-rate} of 15\% on Android-to-OpenHarmony migration, compared to 32\% in Android-to-iOS scenarios~\cite{ReSPlay}, indicating a drop of over 17\%. Similarly, ITeM achieves only 26\% on OpenHarmony, substantially lower than its reported 54\% success-rate in Android-to-Android migration~\cite{ITeM}, corresponding to a decrease of 28\%.  

A similar trend occurs in \underline{step-accuracy}, where ReSPlay achieves 18\%, down from 40\% in Android-to-iOS migration~\cite{ReSPlay}, representing a decrease of 22\%. Similarly, ITeM’s step-accuracy drops to 42\%, compared to 87\% in Android-to-Android migration~\cite{ITeM}, reflecting a decline of 45\%.

\textbf{Result analysis.} We have two observations.

First, although both ReSPlay and ITeM are successfully executed on OpenHarmony after device layer adaptation, their success-rates and step-accuracies remain consistently low across different applications. This means that simple device layer adaptations are insufficient to preserve the effectiveness of migration approaches in the OpenHarmony system. 
For example, existing approaches overlook critical functional cues embedded in the overall layout structure of OpenHarmony system, as they rely on isolated widget analysis paradigms originally optimized for the Android framework.

Second, compared to Android-to-iOS migration, OpenHarmony introduces fundamental differences in system-level execution semantics~\cite{huawei_arkts_guide, huawei_arkui_guide, openharmony_stage_model}. These system-specific characteristics hinder the effective execution of existing GUI test migration approaches. Therefore, bridging this system-level gap requires a detailed investigation of failure modes unique to the OpenHarmony ecosystem.

\Finding{1.1}{Despite implementing device layer adaptation, the evaluated approaches exhibit significantly \textbf{lower effectiveness} in the Android-to-OpenHarmony migration scenario (e.g., ITeM \textbf{26\%}, ReSPlay \textbf{15\%}) compared to Android to other systems, where the success rate drops by 28\% for ITeM and 17\% for ReSPlay. This indicates that effective migration in OpenHarmony is not a \emph{simple interaction adaptation}. Instead, it requires deep deliberation of the \emph{system-specific characteristics inherent to the OpenHarmony architecture}.
}
\vspace{0.5em}
\subsubsection{Execution Cost}
\label{sec:RQ2}

To evaluate the cost of migration, we calculate the \underline{time-per-step} consumed by ReSPlay and ITeM on the ATH Benchmark. Additionally, since ITeM is an LLM-based approach, we also calculate the \underline{token-per-step}.

\textbf{Experimental results.} 
As shown in Table~\ref{tab:rq12}, the time cost of both approaches in the Android-to-OpenHarmony scenario is higher compared with that in the original scenario of ReSPlay and ITeM. 
For ReSPlay, the average latency is 103 seconds per step, 3.5 times higher than that in the original Android scenario (29 seconds).
For ITEM, the average latency is 85 seconds per step, 12 times higher than that in the original Android scenario (7 seconds). The token consumption averages 45,780 per step, which translates to a financial cost of \$0.059 per step using GPT-5, 10,532 tokens (\$0.013) lower than that in the Android scenario (56,312 tokens, \$0.072).

\textbf{Result Analysis.} We analyze that the high time cost for both migration approaches is primarily driven by \textbf{dynamic widgets}, i.e., the GUI widgets with content varying over time or across different executions. To clarify the underlying mechanisms, we analyze the performance of the two approaches separately.

\begin{figure}[t]
\includegraphics[width=0.88\linewidth]{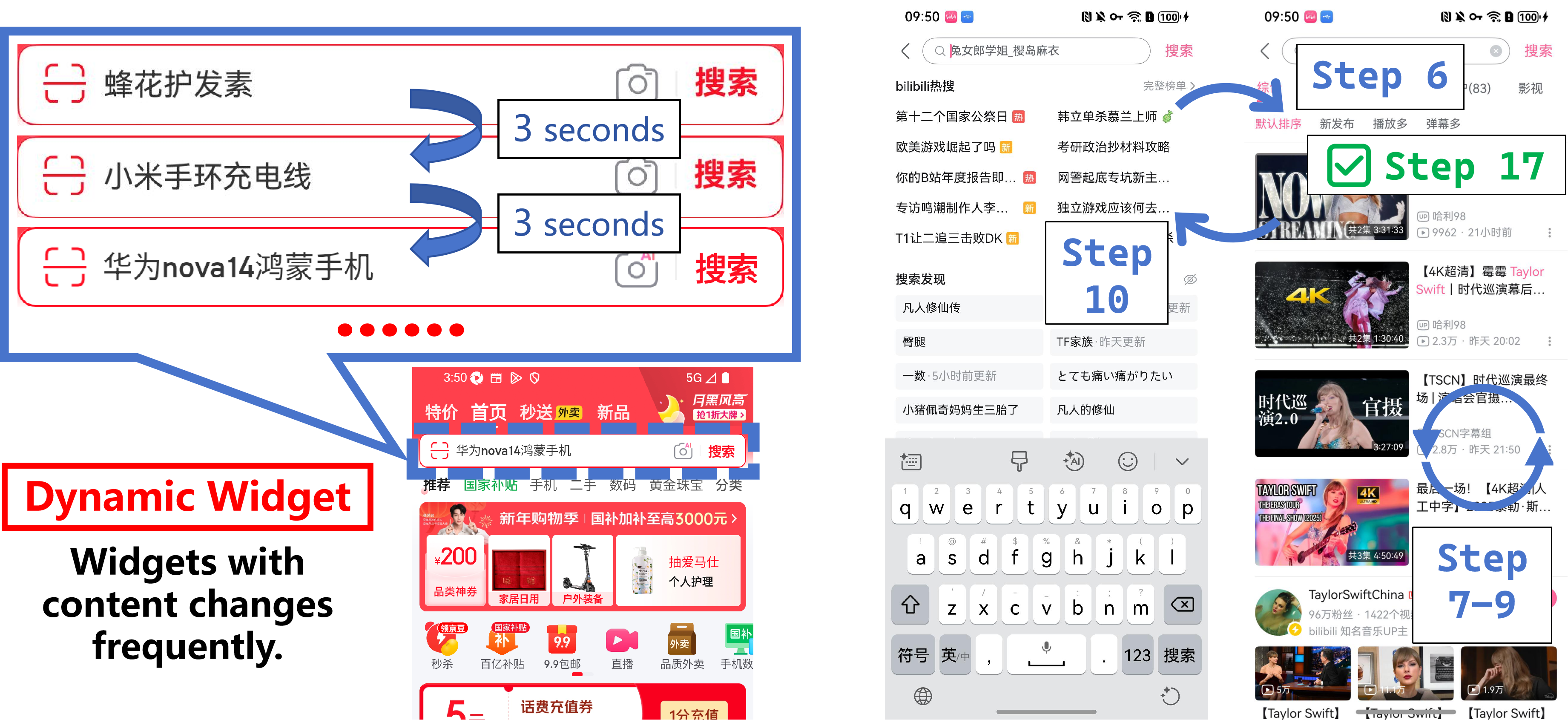}
\caption{Examples of time-consuming cases in ReSPlay and ITeM.}
\label{fig:RQ2}
\vspace{-0.8em}
\end{figure}

For ReSPlay, a representative matching-based approach, it consumes substantially more time in matching targets that contain dynamic widgets. For example, in the \textit{Add to cart} function test, matching the search box takes a huge time cost of more than 10 minutes. As illustrated on the left side of Figure~\ref{fig:RQ2}, the textual content within the search box changes continuously, preventing stable matching based on either textual similarity or visual features. One potential improvement is to incorporate positional information into the matching process, as such dynamic widgets often preserve relatively stable spatial locations across executions in the GUI states.

For ITeM, a representative LLM-based approach, the primary cost contributor is the validation of execution correctness on result pages containing rich dynamic widgets. In the \textit{Search} function test of the Bilibili application~\cite{bilibili}, after performing the final action (i.e., entering search keywords and pressing Enter), the LLM fails to promptly determine whether the intended navigation has completed. Instead, it spends over 8 minutes and consumes more than 200,000 tokens exploring the newly loaded page. As shown on the right side of Figure~\ref{fig:RQ2}, the resulting page is populated with dynamically updated video entries. A possible mitigation strategy is to rely on abstract navigation logic, rather than solely on specific state content.

In summary, dynamic widgets substantially increase the cost of test migration for both matching-based and LLM-based approaches. Mitigating this cost requires migration approaches to integrate multiple GUI information (e.g., texts, state images, and widget location), which helps to better capture the semantics of widgets and GUI states effectively.

\Finding{1.2}{The cost of both ReSPlay and ITeM in the Android-to-OpenHarmony migration scenario is \textbf{higher} compared to that in the Android scenario. \textbf{Dynamic widgets} (e.g., news feeds) cause high consumption in test migration. In matching-based approaches like ReSPlay, dynamic widgets lead to prolonged widget matching; while in LLM-based approaches like ITeM, they result in expensive operation validation. To mitigate the cost, the approaches require \textbf{integrating multiple GUI sources} and \textbf{comprehending widget and state semantics}.}

\subsection{RQ2: Root Causes of Failures}

Driven by the performance (i.e., the effectiveness and efficiency) gaps identified in Section~\ref{sec:RQ1}, we perform an approach-level investigation of the failed cases for both ITeM and ReSPlay to delineate the boundaries of the migration capabilities of the approaches. Specifically, we first present an open coding analysis~\cite{open_coding} that yields three \textbf{categories of causes} in Section~\ref{sec:cause}, followed by a thorough discussion of \textbf{the correlations} between the causes and the design concepts of the frameworks of the approaches in Section~\ref{sec:correlation}.

\begin{table}[t]
\small
  \centering
  \caption{The Influence of Causes on ReSPlay and ITeM}
  \label{tab:rq3}
  \vspace{-0.5em}
  \resizebox{\columnwidth}{!}{
  \begin{tabular}{ccccc}
    \toprule
    \textbf{Level} & \textbf{Causes} & \textbf{ReSPlay} & \textbf{ITeM} & \textbf{Overall} \\
    \midrule
    \multirow{2}{*}{Application Level} & GUI Design Difference & 33\% & 2.5\% & 19\% \\
    & Navigation Difference  & 13\% & 5\% & 9\% \\
    \midrule
    Widget Level & Dynamic Widget & 24\% & 7.5\% & 16\% \\
    \midrule
    \multirow{3}{*}{Program Level} & Erroneous Pruning & 7\%  & 67.5\% & 35\% \\
    & Algorithm Ineffectiveness & 15\% & 17.5\% & 16\% \\
    & Unsupported Interaction & 8\%  & 0\% & 5\% \\
    \bottomrule
  \end{tabular}
  }
\vspace{-1.5em}
\end{table}

\subsubsection{Cause Analysis.} 
\label{sec:cause}
In this section, we present an open coding analysis based on failed cases reproduction and summarize five main causes in three levels.

\textbf{Methodology.} Our analysis is grounded in manual reproduction and the inspection of detailed execution traces. For ITeM, we select a representative run from the three executions for each case, the number of successful steps of which aligns with the result of RQ1 (See Section~\ref{sec:RQ1}). After obtaining the raw results, we employ \textbf{open coding}~\cite{open_coding} method to extract the instructive commonalities of the failures. Specifically, we generate, gather, and categorize the failure situation while observing and comparing the raw execution trace.

\textbf{Result Analysis.} The open coding analysis reveals three core facts. \ding{182}Both approaches successfully record all test cases within the ATH Benchmark. \ding{183}All identified failures occur exclusively during the migration stage. \ding{184}We summarize six underlying causes categorized into three levels (see Table~\ref{tab:rq3}).

\underline{Application-Level Causes} (28\%). These failures stem from application design differences between Android and OpenHarmony environments. Such differences underscore the necessity for migration approaches to bridge the application design gap between Android and OpenHarmony, rather than simple device layer adaptation (see Section~\ref{sec:workflow}).

\begin{enumerate}[label=(\arabic*),leftmargin=*,start=1]
    \item \textit{GUI Design Difference} (19\%): Variations in GUI-specific design elements, such as widget texts, images and widget positions between the Android and OpenHarmony versions of the same applications. These inconsistencies often cause the widget-matching mechanisms of existing approaches to fail due to insufficient similarity scores.
    \item \textit{Navigation Difference} (9\%): Differences in the underlying control logic that governs state transitions and event-triggering mechanisms. For instance, a specific interaction sequence may navigate to a target page on Android but fail to trigger the same transition on OpenHarmony due to differences in application task flows.
\end{enumerate}

\underline{Widget-Level Causes} (16\%). These failures stem from the inherent properties and behaviors of individual GUI widgets, which challenge the widget comprehension capabilities of migration approaches.

\begin{enumerate}[label=(\arabic*),leftmargin=*,start=3]
    \item \textit{Dynamic Widget} (16\%): GUI widgets whose content changes over time or across executions (e.g., news feeds). Such dynamics lead to inconsistencies between the recording stage and the migration stage, increasing the likelihood of widget comprehension and validation failures.
\end{enumerate}
    
\underline{Program-Level Causes} (56\%). These causes originate from the inherent algorithm logic of the migration approaches, revealing that approaches built upon Android-specific assumptions do not generalize well to the OpenHarmony environment.

\begin{enumerate}[label=(\arabic*),leftmargin=*,start=4]
\phantomsection \label{sec:program-causes}

    \item \textit{Erroneous Pruning} (35\%): To lower the execution cost, current approaches design coarse-grained heuristics pruning to reduce the candidate widgets set. However, these pruning lacks comprehension of OpenHarmony-specific widget types and layout structure. This architectural gap leads to the exclusion of essential target widgets from the candidate widgets during the migration stage.
    \item \textit{Ineffective Algorithm} (16\%): Core algorithms of the migration approaches (e.g., widget matching and filtering) fail in special and complex scenarios. For example, the widget matching algorithm of ReSPlay is based on image positioning, which fails to locate a tiny target widget (e.g., a minus shape button) in the cart page containing multiple pictures and information of products.
    \item \textit{Unsupported Interaction} (5\%): Current migration approaches are restricted to a subset of interactions. Certain interaction types (e.g., \texttt{double-click} and \texttt{long-click}) are not supported by these approaches.
\end{enumerate}

\subsubsection{Correlations between Causes and Approaches.} 
\label{sec:correlation}
Table~\ref{tab:rq3} illustrates the different impact of these causes on ReSPlay and ITeM. ReSPlay suffers primarily from application-level causes (46\%) while ITeM suffers mainly from program level causes (85\%). This difference reflects the divergence of the \textit{design concepts} of the approaches, summarized as follows.

\begin{itemize}[leftmargin=*]
    \item \underline{Application-level causes} significantly impact ReSPlay (46\%) as it strongly assumes the textual and visual consistency of cross-system application design. Based on this assumption, ReSPlay's validation mechanism hinges on a predefined similarity threshold for comparison, even minor system-specific design variations of applications may cause the similarity score to fall below this threshold, triggering a false failure. In contrast, ITeM is less sensitive to application-related differences due to its LLM-based semantic reasoning capability.
    \item \underline{Widget-level causes} affects both approaches (24\% in ReSPlay and 8\% in ITeM), indicating that traditional widget comprehension challenge observed in Android scenario~\cite{LIRAT, MAPIT} persists in the OpenHarmony scenario. Both approaches attempt to extract the functional semantics of the widget from the textual features. However, as the textual content changes frequently in dynamic widget, the inference of the approaches is not reliable. This emphasizes making use of the clues hidden within the overall layout structure.
    \item \underline{Program-level causes} seriously hinder ITeM (85\%) as it relies on a significant characteristic of Android UI hierarchy that the functional semantics strongly binds to a specific widget. Therefore, ITeM only selects clickable top-layer widgets as candidates. As the characteristic is no longer suitable for the unique structural paradigms of OpenHarmony (i.e., ArkUI framework), ITeM incorrectly excludes the target widgets frequently.
\end{itemize}

\Finding{2}{The causes of migration failure are multi-faceted, spanning \textbf{application design difference}, \textbf{challenging widget comprehension}, and \textbf{program inherent limitation}. While widget-level issues are intrinsic to GUI test migration in both Android and OpenHarmony environments, application-level and program-level causes are significantly more pronounced in OpenHarmony. This disparity emphasizes that to improve migration effectiveness, research must transfer \emph{superficial adaptations} to \emph{fundamentally reconstruct Android-oriented algorithms}. In particular, it is highlighted to align the widget matching and filtering heuristics with the unique structural paradigms of OpenHarmony.
}

\subsection{RQ3: System Characteristic Tracing}
\label{sec:RQ3}
To further understanding the conflict between the design concepts of the approaches (i.e., ReSPlay and ITeM) and the OpenHarmony system, we conduct a system-oriented charateristics investigation and a quantified analysis.

\subsubsection{Methodology} First, we summarize the main characteristics that distinguish OpenHarmony from Android through a \textit{literature analysis} of multiple related works~\cite{haptest,haprepair,roadmap,llmrr,hmtest,typescript2arkts} and official Openharmony documents~\cite{container_comp,openharmony_stage_model,customization_capabilities,custom_composition,huawei_arkts_guide,huawei_arkui_guide,huawei_arkui_stage}. Second, we conduct a quantified relational analysis between the failed cases (the same as RQ2) and the observed characteristics to identify their extent of influence.

\subsubsection{Result Analysis} 

Table~\ref{tab:rq4} presents the proportion of failed test cases associated with different factors. Overall, 28\% of the failed cases are mainly caused by non-system factors, such as dynamic widgets and unsupported interactions. In contrast, the remaining 72\% are system-related and can be traced to characteristics specific to OpenHarmony. These system-related factors can be further grouped into two aspects, namely \emph{technical architecture differences} and \emph{ecosystem characteristics}, which together comprise five characteristic categories. In the following, we analyze these five categories in detail and discuss how they contribute to migration failures.

\begin{table}[t]
\small
  \centering
  \caption{The Failed Test Cases Related to the Characteristics}
  \label{tab:rq4}
  \vspace{-0.5em}
  \resizebox{\columnwidth}{!}{
  \begin{tabular}{cccc}
    \toprule
    \textbf{Perspectives} & \textbf{Characteristics} & \textbf{Count} & \textbf{Proportion} \\
    \midrule
    Technical & UI framework     & 23 & 27\% \\
    Architecture & Page management  & 6  & 7\%  \\
    Differences & Widget design    & 10 & 12\% \\
    \midrule
    Ecosystem & Evolving stage   & 21 & 24\% \\
    Characteristics & Community resources & 2 & 2\%  \\
    \midrule
    Others & Non-system factor & 24 & 28\% \\
    \midrule
    Overall & & 86 & 100\% \\
    \bottomrule
  \end{tabular}
  }
\vspace{-1.5em}
\end{table}

\textbf{Technical Architecture Differences (46\%).} The architectural differences between OpenHarmony and Android include different UI framework, application page management, and widget design, reflecting a \textit{holistic design philosophy} (i.e., integral UI design and compositional widget design) in OpenHarmony distinguish from Android.

\underline{UI framework (27\%).} OpenHarmony adopts a declarative UI framework ArkUI~\cite{haptest} (see Section~\ref{sec:HMbackground}), a paradigm shift from Android's imperative XML-based UI model. In ArkUI, layout structures are generated dynamically, and interactive functional logic is typically attached to container widgets~\cite{container_comp} that encompass multiple widgets. The emphasis of structural information in integral UI design reflects OpenHarmony's \textit{holistic design philosophy}. This architectural difference directly contributes to \textit{Erroneous Pruning} (recall Section~\ref{sec:cause}) as both evaluated approaches only focus on the top-level widgets, instead of the whole structure, which does not align with the holistic and dynamic layout structure.

\underline{Application page management (7\%).} OpenHarmony employs the Stage Model~\cite{openharmony_stage_model} (see Section~\ref{sec:HMbackground}), which fundamentally differs from Android's Activity-centric model. In the Stage Model, a single \texttt{UIAbility} manages multiple ArkUI pages~\cite{hmtest}, whereas Android typically binds one Activity to a single XML layout file. This structural difference complicates execution-state validation for ITeM  (recall Section~\ref{sec:RQ2}) 
, as it does not distinguish between the visible and invisible widgets coexisting within the same ArkUI page, misleading to misjudgments when validating operation outcomes.

\underline{Widget design (12\%).} OpenHarmony encourages extensive use of custom widgets and compositional design, promoting service-oriented and holistic widget groups~\cite{customization_capabilities, custom_composition}, i.e., design custom widget groups to implement specific function for the service. This design philosophy renders ReSPlay's reliance on matching text from widgets on the top layer of the hierarchy ineffective, resulting in incorrect widget alignment. Similarly, ITeM's isolated and type-specific widget analysis leads to \textit{Erroneous Pruning} (recall Section~\ref{sec:cause}). This highlights that effective migration on OpenHarmony requires a more service-oriented and holistic GUI comprehension through making use of the clues hidden within the overall layout structure.

\textbf{Ecosystem Characteristics (26\%).} Beyond architectural differences, the nascent evolving stage and scarcity of resources of the OpenHarmony ecosystem introduce additional challenges, which surface as state inconsistency and reasoning limitations in migration approaches.

\underline{Evolving stage (24\%).} The HarmonyOS NEXT ecosystem remains relatively nascent, rapidly evolving and largely isolated from Android~\cite{roadmap}. This results in applications with redesigned GUIs and rapid iteration cycles to leverage exclusive features. This explains the \textit{application-level causes} of failures (recall Section~\ref{sec:cause}) , which are caused by the approaches' heavy reliance on cross-system GUI similarity. In practice, assumptions of GUI stability and visual consistency are frequently violated on OpenHarmony, instead dynamic and function-focused matching is emphasized.

\underline{Community resources (2\%).} The limited availability of open-sourced frameworks and public repositories for ArkTS (the official high-level language for OpenHarmony application development~\cite{huawei_arkts_guide, haprepair, typescript2arkts, roadmap}) results in insufficient training data for LLM-based approaches. Therefore, LLMs struggle to accurately reason about OpenHarmony-specific execution semantics, especially the hidden navigation logic. This limitation is reflected in ITeM on the application level (see Section ~\ref{sec:cause}), where critical widgets are excluded during LLM's reasoning. 

\Finding{3}{The OpenHarmony's inherent characteristics summarized from literature sources do contribute to the failure in statistics (72\%). Its technical architecture differences (46\%) reflect a \textit{holistic design philosophy} and mainly results in \textit{Erroneous Pruning} by the declarative \textbf{UI framework} with dynamic layouts and container-based logic and the custom, service-oriented \textbf{widget design}. While its ecosystem characteristics (26\%)  contribute to \textit{application-level causes} because the ecosystem's nascent, rapidly textbf{evolving stage}  violates assumptions of GUI stability. These characteristics collectively challenge Android-oriented algorithms for migration approaches.}
\section{ITEM-HM}

Based on the identified OpenHarmony-specific characteristics that hinder test migration, we design and implement an enhancement approach based on ITeM, referred to as \textbf{ITeM-HM}. Specifically, Section~\ref{sec:enhance-design} presents the design of the enhancements, which focus on system-specific optimization. Subsequently, Section~\ref{sec:RQ5} evaluates the effectiveness of ITeM-HM in Android-to-OpenHarmony test migration and analyzes the impact of these enhancements on the final results. Finally, Section ~\ref{sec:ablation} conducts an ablation study to evaluate the contributions of each enhancement technique in ITeM-HM.

\subsection{Design of ITeM-HM}
\label{sec:enhance-design}
We select ITeM as the subject for enhancement due to its superior baseline performance and its significant constraints from OpenHarmony-specific characteristics, particularly its \textit{holistic design philosophy} (see Section~\ref{sec:RQ3}). We redesign three core components (See Section~\ref{sec:approaches}) of ITeM to address the specific failure patterns observed in RQ2 (see Section~\ref{sec:RQ2}) and RQ3 (see Section~\ref{sec:RQ3}).

\begin{figure}[t!]
    \centering
    \includegraphics[width=0.92\linewidth]{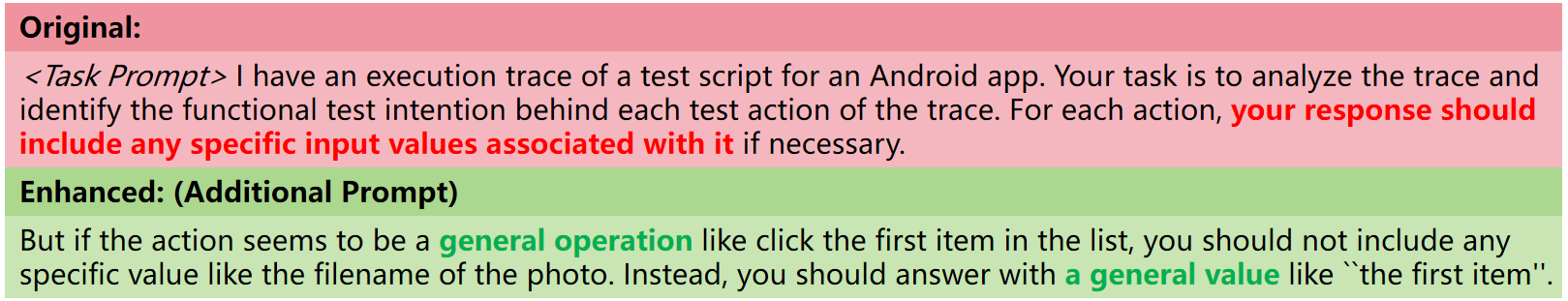}
    \vspace{-0.4em}
    \caption{The enhanced task prompt to abstract the general test logic.}
    \label{fig:RQ5-task-prompt}
\vspace{-0.3em}
\end{figure}

The \textbf{test intention extraction} component use LLM to analysis the execution trace of the Android test case and summarized the intention of each step into a sequence. As shown in Figure~\ref{fig:RQ5-task-prompt}, in the original ITeM, instructions to preserve ``specific input values'' often cause the LLM to overfit instance-specific details (e.g., a particular search result like ``Taylor Swift’s Year-End Message''), leading to migration failures when such detail information differs on OpenHarmony. 

To avoid overfitting, we \textbf{abstract the general test logic} from the record stage, removing the application and system-specific information. Specifically, we refine the task prompt to guide the LLM towards extracting higher-level, generalized functional test logic. As shown in Figure~\ref{fig:RQ5-task-prompt}, the enhanced algorithm in ITeM-HM explicitly instructs the LLM to \textbf{infer general operations} (e.g., ``click the first item in the list''), thereby improving the robustness and transferability of the extracted test logic across systems.

\begin{figure}[t!]
    \centering
    \includegraphics[width=0.8\linewidth]{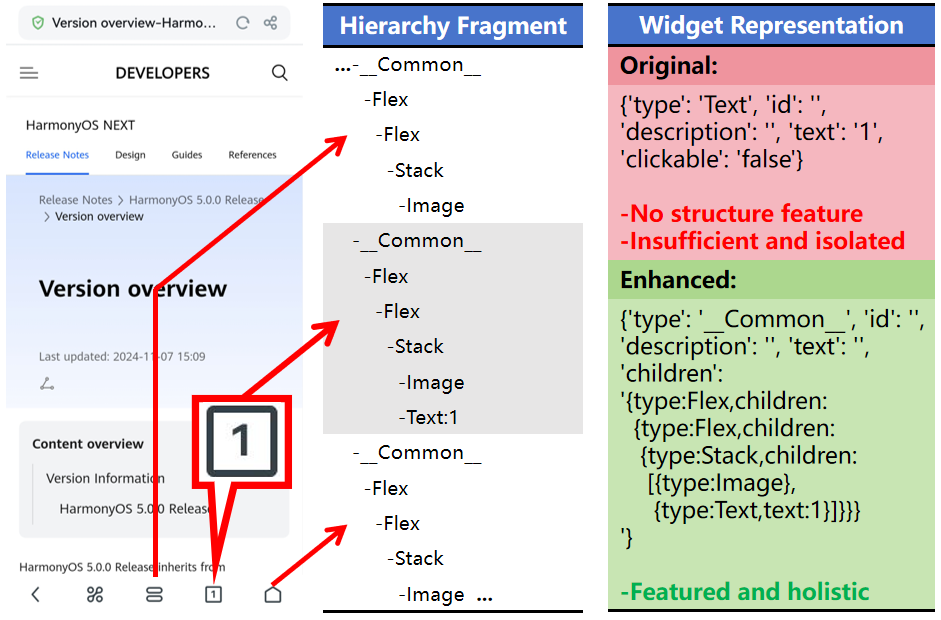}
    \vspace{-0.4em}
    \caption{Example of widget representation of \textit{functional widget group} (the tab button in UC Browser).}
    \label{fig:RQ5-widgets-prompt}
\vspace{-0.5em}
\end{figure}

The \textbf{GUI content extraction} component preliminarily filters candidate widgets and represents them with a specific structure to prompt LLM. The original ITeM only selects top-layer widgets and represents them with their limited and nonspecific individual attributes (e.g., \texttt{type}, \texttt{description}, \texttt{text}, and \texttt{clickable}).

To align with OpenHarmony's UI framework and service-oriented and holistic widget groups~\cite{customization_capabilities, custom_composition} (Recall Section~\ref{sec:RQ3}), we \textbf{model functional widget groups}. Specifically, we revise the representation of widget in two aspects.

\begin{itemize}[leftmargin=*]
    \item First, we generalize the interactivity analysis heuristic. Instead of relying solely on the Android-centric \texttt{clickable} attribute and specific widget types, we also consider widgets whose type field contains functional keywords (e.g., button, edit, input) to include custom widgets. 

    \item Second, we introduce the concept of functional widget groups. In ArkUI, interactive behavior is often designed at high-level containers (see Section~\ref{sec:RQ3}) (e.g., \texttt{Common}, \texttt{Stack}), while their containing widgets (e.g., \texttt{Image}, \texttt{Text}) provide the visual and semantic details (see the middle of Figure~\ref{fig:RQ5-widgets-prompt}). Therefore, we aggregate the information of such child widgets into a \texttt{children} attribute of the parent container. This grouping strategy makes effective use of the clues hidden in the layout structure and enables a more faithful representation of the actual functional logic of the OpenHarmony GUIs, as shown in Figure~\ref{fig:RQ5-widgets-prompt}.
\end{itemize}

The \textbf{action execution} component executes LLM's actions in dynamic reasoning during migration. In the original ITeM, LLM selects a target widget and an operation from a predefined set. The component relocates the position of the widget in stage with its attributes and executes the operation. This \textit{widget relocation mechanism} frequently misaligns LLM intents with fixed widget, especially when its textual features are limited, hindering exploration of alternative interactions.

To take full advantage of the exploration of LLM, we \textbf{enhance the robustness and expressiveness of the interaction layer} through two key enhancements.

\begin{itemize}[leftmargin=*]
    \item First, we remove the widget relocation mechanism and replace it with a \textbf{bounds-caching strategy}, which assigns unique identifiers to all interactive widgets and caches their bounds. When the LLM specifies a target widget by its identifier, the corresponding action is executed directly using the cashed bounds, avoiding error-prone relocation. 

    \item Second, we extend the supported interactions on OpenHarmony, including text input without navigation and more gesture operations (e.g., \texttt{double-click}, \texttt{long-click}, and multi-directional \texttt{swipes}), enabling ITeM-HM to handle richer OpenHarmony interaction patterns.
\end{itemize}

\subsection{RQ4: Performance of ITeM-HM}
\label{sec:RQ5}

To assess the capability of ITeM-HM, we evaluate its effectiveness and efficiency using the ATH Benchmark. This evaluation focuses on measuring the performance gains achieved through our OpenHarmony-specific enhancements, compared to the original ITeM framework. The metrics are the same as in the RQ1, which are \underline{success-rate}, \underline{step-accuracy}, \underline{time-per-step}, and \underline{tokens-per-step} (see Section~\ref{sec:metrics}). Similar with Section~\ref{sec:RQ1}, given the non-determinism of LLMs, we conduct three independent runs and adopt the median results.

\begin{table}[t]
\centering
\small
\caption{Comparison of ITeM and ITeM-HM}
\label{tab:rq5_results}
\vspace{-0.5em}
\resizebox{\columnwidth}{!}{
\begin{tabular}{lccc}
\toprule
\textbf{Metrics} & \textbf{ITeM} & \textbf{ITeM-HM} & \textbf{Improvement} \\
\midrule
Success-rate & 26\% & 81\% & 214\%$\uparrow$ \\
Step-accuracy & 42\% & 86\% & 107\%$\uparrow$ \\
Time(s/step) & 85 & 47 & 44\%$\downarrow$  \\
Token(tokens/step) & 45,780 & 19,717 & 57\%$\downarrow$ \\
\bottomrule
\end{tabular}}
\vspace{-1.5em}
\end{table}

\textbf{Experimental results.} For effectievnes, ITeM-HM achieves a success-rate of 81\% and a step-accuracy of 86\% (see Table~\ref{tab:rq5_results}). Compared to ITeM (26\% success-rate and 42\% step-accuracy, see Section ~\ref{sec:RQ1}), ITeM-HM improves the success-rate by \textbf{214\%} and the step-accuracy by \textbf{107\%}.

For effiency, the average time and token cost of ITeM-HM are reduced to 41 seconds and 20,255 per step, which translates to a financial cost of \$0.037 per step using GPT-5, compared to ITeM (83 seconds, 56,312 tokens, \$0.099).

\textbf{Result analysis.} ITeM-HM effectively addresses the dominant cause, which is rooted in OpenHarmony's holistic design philosophy (recall Section~\ref{sec:RQ3}). The proposed enhancements solve the major failure cause (i.e., \textit{Erroneous Pruning} in Section~\ref{sec:cause}). Details are as follows.

Among the enhancements, the introduction of \textit{functional widget groups} (recall Section~\ref{sec:enhance-design}) plays a critical role, resolving 10 out of 14 cases of \textit{Erroneous Pruning}. This confirms that providing the LLM with a functionality-focused interactive widget group is crucial for corresponding with the holistic design philosophy of OpenHarmony.

In terms of operational consumption, ITeM-HM significantly reduces per-step costs. The reasons are that although the representation of individual components becomes richer, the \textit{additional GUI context} and \textit{structural information} significantly reduce the need for costly dynamic exploration caused by insufficient semantic cues. As a result, the number of interactions required between the LLM and the system to complete a migration task is substantially reduced. This leads to a notable decrease in both time cost and token consumption in the overall process.

\Finding{4.1}{\underline{For effectiveness}, ITeM-HM achieves \textbf{significantly higher} effectiveness, increasing the success-rate from 26\% (ITeM) to \textbf{81\%} (ITeM-HM), with the improvement of \textbf{214\%}. This remarkable enhancement is primarily due to designing the \emph{functional widget groups} to mitigate the \textit{Erroneous Pruning} based on the holistic design philosophy of OpenHarmony. \underline{For efficiency}, ITeM-HM significantly \textbf{reduces per-step costs} to 41 seconds and 20225 tokens (\$0.037), compared to the baseline ITeM (83 seconds, 68860 tokens, \$0.121). This reduction is attributed to the \textit{additional GUI context information}, which provides more precise semantic guidance and significantly decreases the number of LLM conversations for dynamic exploration.}
\vspace{0.4em}

Despite the improvement, a persistent challenge remains as migration failures caused by \textit{image-type widgets}, which lack the explicit textual semantics necessary for current LLM-based approaches to accurately interpret their functional intent. A potential solution lies in the integration of vision-based comprehension with traditional text-based analysis to bridge this semantic gap.

\Finding{4.2}{The persistent challenge of \textbf{image-type widgets} underscores the inherent limitations of traditional text-only LLMs and highlights the critical necessity for \textbf{multi-modal information integration} to achieve robust GUI test migration.}

\subsection{RQ5: Ablation Study}
\label{sec:ablation}
To evaluate the specific contributions of each enhancement in ITeM-HM, we conduct an ablation study by progressively incorporating our proposed components. The results are shown in Table~\ref{tab:rq5_ablation} and summarized as follows:

\begin{table}[t]
\centering
\small
\caption{Ablation Results}
\label{tab:rq5_ablation}
\vspace{-0.5em}
\resizebox{\columnwidth}{!}{
\begin{tabular}{lcccc}
\toprule
\textbf{Metrics} & \textbf{ITeM} & \textbf{ITeM-prompt-enhanced} & \textbf{ITeM-GUI-enhanced} & \textbf{ITeM-HM} \\
\midrule
Success-rate & 26\% & 26\% & 74\% & 81\% \\
Step-accuracy & 42\% & 44\% & 80\% & 86\% \\
\bottomrule
\end{tabular}
}
\vspace{-1.5em}
\end{table}

\begin{itemize}[leftmargin=*]
    \item Original ITeM (with only device layer adaptation): achieves a success-rate of 26\% and a step-accuracy of 42\%.
    \item ITeM-prompt-enhanced (with enhanced test intention extraction): maintains a similar success-rate of 26\%, while improving step-accuracy to 44\%.
    \item ITeM-GUI-enhanced (further incorporating enhanced GUI content extraction): significantly improves performance to a success-rate of 74\% and a step-accuracy of 80\%.
    \item ITeM-HM (additionally incorporating enhanced action execution): further improves performance to a success-rate of 81\% and a step-accuracy of 86\%.
    
\end{itemize}

\Finding{5}{The ablation study confirms that the enhancement of all the three components (i.e., test intention extraction, GUI content extraction, and action execution) contribute complementary improvements, while the major improvement (48\% for success-rate) stems from the \textbf{GUI content extraction enhancement}.}
\section{Threat To Validity}
\label{sec:threattovalidity}

\textit{Internal Validity} mainly concerns potential biases introduced during experimental design, tool adaptation, and test execution. To mitigate such threats, we adopt a unified and controlled experimental workflow across all evaluated applications and testing techniques. The adaptations to GUI test migration techniques are strictly limited to enabling their execution in the OpenHarmony environment, without modifying their core testing logic. In addition, all experiments are conducted under the same hardware configuration and OpenHarmony version to minimize the impact of environmental differences. These measures help reduce confounding factors and improve the reliability of the experimental results.

\textit{External Validity} is threatened by the representativeness of the selected subjects. To alleviate this threat, we carefully select 36 popular applications with an average download of 9 billion, covering six common application categories, which helps improve the diversity and representativeness of the dataset. As for the limitation of functionalities, we prioritized primary functionalities, which highlighted in official store descriptions and identified as "core" in established benchmarks~\cite{craftdroid,RIDA}. Moreover, our study evaluates two representative approaches for GUI test migration, ensuring our insights are broadly applicable to the current research landscape.

\textit{Construction Validity} primarily includes the human bias of selection and the potential of data leakage. First, regarding human bias during the manual selection of applications and functionalities, we align our criteria with established methodological benchmarks (e.g., CraftDroid~\cite{craftdroid}) and prioritize core workflows highlighted in official Google Play Store descriptions to ensure our test cases reflect objective user priorities. Second, we consider the threat of data leakage arising from the integration of LLM within ITeM. Since LLM-based approaches may rely on knowledge or patterns seen during previous execution cycles, there is a risk that the ground-truth information could be leaked to the model, potentially leading to unreliable performance metrics.
\section{Conclusion}
In this paper, we present the first empirical study of GUI test migration from Android to OpenHarmony. We construct a benchmark dataset of 36 representative applications (forming 18 cross-system pairs) with an average downloads of over 9 billion and 108 manually designed test cases, and systematically evaluate two representative test migration techniques, ITeM and ReSPlay. After applying necessary device-level adaptations to enable these techniques execution in OpenHarmony, we analyze these techniques across three dimensions: effectiveness and efficiency, failure causes, and system-specific characteristics. Based on the analysis, we propose an OpenHarmony-oriented enhancement ITeM-HM, which significantly enhances migration effectiveness. Furthermore, we conduct an ablation study of each enhancement of ITeM-HM, which demonstrates that the major improvement stems from the GUI content extraction enhancement.

\section{Data Availability}
We open-source our dataset benchmark and approach code on \url{https://anonymous.4open.science/r/GUI-Test-Migration/}.


\end{document}